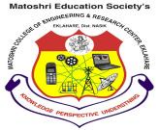
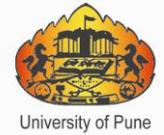

# Authentication Mechanism for Resistance to Password Stealing and Reuse Attack

Sharayu A. Aghav and RajneeshKaur Bedi

**Abstract** Considering computer systems, security is the major concern with usability. Security policies need to be developed to protect information from unauthorized access. Passwords and secrete codes used between users and information systems for secure user authentication with the system. Playing a vital role in security, easily guessed passwords are links to vulnerability. They allow invader to put system resources significantly closer to access them, other accounts on nearby machines and possibly even administrative privileges with different threats and vulnerabilities (e.g., phishing, key logging and malwares). The purpose of this system is to introduce the concept and methodology which helps organization and users to implement stronger password policies. This paper studies a password stealing and reuse issues of password based authentication systems. Techniques and concepts of authentication are discussed which gives rise to a novel approach of two-factor authentication. Avoiding password reuse is a crucial issue in information systems which can at some extent contribute to password stealing issue also. In the proposed system, each participating website possesses a user's unique phone number, telecommunication services in registration and recovery phases and a long-term password used to generate one-time password for each login session on all websites.

*Index Terms*— **Authentication, Network security, One-time password, Phishing.**

## I. INTRODUCTION

In the current public networks, since most of the activities are available on internet, user authentication is the most important part as far as security is concerned. Internet connected PCs are not 'safe' anywhere, most of the user computers are infected with one or more forms of spyware or malware. Software key loggers are installed on a user PC along with common malware and spyware [8]. An increasing number of phishing sites also install key loggers on user PCs, even when users do not explicitly download or click any link on those sites [9]. Phishing is a malicious activity whereby an intruder (phisher) tries to trick internet users into providing confidential information (Dhamija et al., 2006). It is a severe problem because phishers can steal sensitive information, such as user's bank account details, social security numbers and credit card numbers by masquerading as a trustworthy entity in an electronic communication. Although many more secure (typically also more complex and costly) authentication protocols have been proposed, due to the ease of deployment and usability issue; Passwords is ubiquitous technique for online authentication even in such an unsecure environment.

Computer users are asked to create, keep and remind an increasing number of passwords for host accounts, e-commerce sites, email servers and online financial services. Usually, password-based user authentication can resist brute force and dictionary attacks if users select strong passwords to provide sufficient entropy. Unfortunately, the password entropy that users can easily memorize seems inadequate to store unique and secure passwords for all these accounts causes users be likely to reuse passwords across different websites [6], [7]. In 2007, Florencio and Herley [11] specified that a user reuses a password across 3.9 various websites regularly. Password reuse is the reason for users to lose confidential information stored in various websites if a hacker compromises one of their passwords. This attack is referred as the password reuse attack. Hence, passwords are a primary target of attackers for economically-motivated exploits including those targeting online bank accounts and identity theft.

Some experts focus on three-factor authentication instead of only password-based authentication. Three-factor authentication depends on what you know (e.g., password), what you have (e.g., token), and who you are (e.g., biometric). To get authenticated, the user need to input a password and provide a pass code generated by the token (e.g., RSA SecureID [12]), and scan his biometric features (e.g., fingerprint or pupil). Three-factor authentication is a complete defense mechanism in opposition to password stealing attacks, but it needs proportionally high cost [13]. Consequently, knowing the limitations of single-factor (password-based) authentication, two-factor authentication is more possible and reliable even than three-factor authentication. Even though many banks implement two-factor authentication, it still experiences the drawbacks, such as the password reuse attack. Users have to remember an additional four-digit PIN code to work mutually with the token, for example RSA SecureID. We also have to consider, users can easily forget to carry the token. So, proposed system aims to improve the security and usability of two-factor authentication mechanism, as a result opposes most of the possible attacks and has a fewer requirements when compared to other systems.

## II. LITERATURE SURVEY

B. Parno *et al.* [2] proposed a system that guards the secrecy and integrity of a user's existing online accounts so that attacks are no more successful than pre-internet scams (e.g., an attacker may still be able to access a user's account by weakening a company insider). System is based on the observation that users should be authenticated using an additional authenticator that they cannot readily reveal to malicious parties, authenticator can take a form of a cell phone, PDA or even a smart watch; in this method a cell phone is used. This scheme establishes the additional authenticator on a trusted device, for that an attacker must compromise the device and gain the user's password to access the user's account. Users cannot readily reveal the authenticator on the cell phone to a third party, and servers will decline to act on instruction received from someone claiming to be a particular user without presenting the proper authenticator. This technique is one of the first systems to avoid active Man-in-the-Middle attack. In addition, the use of the cell phone reduces the effect of hijacked browser windows and facilitates user convenience. It is assumed that the user can establish a secure connection between their cell phone and browser and the cell phone itself has not been compromised. Again, Phoolproof is yet vulnerable to the password reuse problem and require physical contacts to ensure that account setup is secure.

M. Mannan *et al.* [4] stated that the primary goal of MP-Auth is to protect user passwords from malware and phishing websites and to provide transaction integrity. Here assumption made that a bank's "correct" public key is available to users and mobile devices are malware-free. SSL connection establishment between bank server and a browser on a PC by using a bank's SSL certificate (as per common current practice). The browser may be tricked to go to a spoofed website or have a wrong SSL certificate of the bank or the verifying certificating authority. Denial-of-service (DoS) attack not addressed. Despite of that, MP-Auth suffers from password reuse vulnerability. An invader can compromise a weak server, e.g., a server without security patches to gain a victim's password and use it to get his access rights of different websites. Assumption made in MP-Auth is that account and password setup is secure. Users should setup an account and password by personal communication with the bank to initialize their account or bank will provide passwords through postal service.

Justin Martineau *et al.* [5] proposed a system in which the user's home/office computer works as a trusted proxy that authenticates the user to the service; user is trying to access from the un-trusted machine. The user's authentication details, such as usernames and passwords for sites and services are kept and protected on the proxy. When a user requests access to a secure website from the un-trusted machine, the proxy will fill in the authentication details on behalf of the user, provided the un-trusted machine provides a valid response to the proxy's one time password challenge. The user's mobile device computes correct response to this challenge; it is initialized with a onetime password program in union with the proxy. Both the proxy and the mobile device are in sync with respect to the one time password generation. If the response to the challenge matches the value computed by the proxy then the proxy supplies the proper authentication details to the trusted web site. Thus users can authenticate to internet services without disclosing any long time passwords to the untrusted machine. The scheme prevents the un-trusted machine from capturing and reusing the user's passwords and other authentication data but does not guarantee that the user's password was used for authenticating to the user specified website, e.g., the user could have sent the one time password for www.amazon.com, while the un-trusted machine could have modified the destination address to www.buy.com and still sent the valid one time password. The proxy wouldn't be able to detect this modification and would supply the authentication details for www.buy.com. In order to thwart this attack, the one-time password used for authentication must be dependent upon both the user's secret key and the site the user is trying to access. This can be accomplished using AES. System faces a problem of the overhead of site encryption for secure site identification.

oPass [1] provides users with a mechanism to authenticate from an un-trusted machine to an Internet service without revealing confidential user information which uses a user's cell phone and short message service (SMS) to prevent password stealing and password reuse attacks. We can say that it is difficult to prevent password reuse attacks from any scheme where the users need to recall something. We also state that reason behind password stealing is when users type passwords to untrusted public computers. Therefore, aim of oPass is free users from having to remember or type any passwords into malicious computers for authentication. Different from common user authentication, oPass involves a new element, the cell phone which is used to generate one-time passwords and a new communication medium, SMS which is used to transmit authentication messages. The telecommunication service provider (TSP) plays a role in the registration and recovery phases. The TSP acts as a link between subscribers and web servers. It offers a service for subscribers to execute the registration and recovery process with each web service, e.g., a subscriber inputs her id $ID_u$ and a web server's id $ID_s$ to start the registration phase. Afterwards, the TSP forwards the request and the user's phone number to the related web server based on the received $ID_s$. Still, oPass is vulnerable to the attacks possible due to the SSL channel and requires more secure connection with low communication overhead.

*OTHER RELATED WORK.* Also, the other parallel researches have given different methods to avoid phishing attacks. SessionMagnifier is a technique that facilitates extended browser on a mobile device along with a regular browser on a public computer that work together to secure a web session [14]. SessionMagnifier detaches user access to very sensitive interactions (online banking or payment) from the regular interactions (web surfing or photo viewing). For very sensitive interactions, the information is always sent to the extended browser on the users' mobile device for further verification. Another approach is adopting TPM (Trusted Platform Module). McCune et al. developed a system which is known as a bump in ether (BitE) adopted TPM [15]. BitE facilitates, all user inputs will be protected under an encrypted tunnel, a secure connection between the mobile device and an application running on a TPM based untrustworthy computer.

Garriss et al. designed another system based on TPM and virtual machine (VM) technologies [16], to guarantee trustworthiness of public kiosks for secure computing.

### III. IMPLEMENTATION DETAILS

The proposed system is novel architecture for a user authentication to thwart phishing and password reusing attacks. The purpose of protocol is to avoid users from typing their memorized passwords into public kiosks. By adopting one-time passwords, password information is no longer useful. A one-time password is expired when the user finishes the current session. Different from using internet channels, leverages SMS and user's cell phones to prevent password stealing attacks. We believe SMS is a secure and suitable medium to pass on important information between cell phones and websites. Based on SMS, a user identity is authenticated by websites without inputting any passwords to untrusted kiosks. User password is only used to limit access on the user's cell phone. In system, each user simply memorizes a long-term password to access her cell phone. The long-term password is used to guard the information on the cell phone from a theft.

The assumptions made in system are as follows.
1) Every web server owns a unique phone number. Through a SMS channel, users can interact with each website using the phone number.
2) The telecommunication service provider plays a role in the registration and recovery phases. The TSP module is a link between subscribers and web servers which resides at server only. It offers a service for subscribers to perform the registration and recovery progress with each web service e.g., a subscriber inputs her id $ID_u$ and a web server's id $ID_s$ to execute the registration phase. Afterwards, the TSP module sends the request and the subscriber's phone number to the related web server based on the received $ID_u$.
3) Subscriber's (i.e., users) establishes connection to the server with TSP module through 3G connections.
4) If a user loses her cell phone, he can inform his service provider (TSP) to disable her misplaced SIM card and keeps a new card with the same phone number. Hence, the user finishes the recovery phase.

#### A. System Overview

Unlike general web logins, system makes use of a user's cell phone as an authentication token and SMS as a secure communicating medium. In the *registration* phase, a user starts the program to register his new account on the website, she desires to access later. Contrasting with conventional registration, the server asks for the user's account id and phone number, in spite of password. After filling out the registration form, the program asks the user to enter a long-term password. A chain of one-time passwords for next logins on a target server is generated by using long-term password. Afterwards, the program involuntarily sends a registration SMS to the server to finish the registration procedure. The registration SMS is encrypted to provide data confidentiality. A *recovery* phase is designed to fix problems in some circumstances, such as losing one's cell phone.

Contrasting with common cases, *login* process in system does not require users to type passwords into an untrusted web browser. The user name is the only input information to the browser. Further, the user starts the program on her phone and enters the long-term password; the program will produce a one-time password and send a login SMS safely to the server. The login SMS is encrypted by using the one-time password.

At last, the cell phone receives a response message from the server and shows a success message on her screen if the server is able to confirm her identity. The message is used to guarantee that the website is an authorized website and not a phishing one. We follow same system overview as that of the overview presented in oPass [1]. Procedure of each phase is presented as below.

#### B. Registration Phase

Fig.1 describes registration phase. This phase aims to allow a user and a server to agree on a shared secret to authenticate subsequent logins for this user. The user opens the program on her cell phone. She inputs $ID_u$ (account id) and $ID_s$ (generally the website URL or domain name) to the program. The mobile application then sends $ID_u$ and $ID_s$ to the telecommunication service provider (TSP) which is a module at server only; through a 3G connection to ask for a registration request. The TSP module got the $ID_u$ and the $ID_s$, it is able to trace the user's phone number $T_u$ based on user's SIM card. The TSP module acts as the third-party to distribute a shared secrete key $K_{sd}$ between the user and the server. Encrypt the registration SMS with AES-CBC by using the shared secrete key $K_{sd}$. Then the TSP sends $ID_u$, $T_u$ and $K_{sd}$ to the related server. Server will produce the related information for this particular account and reply with a response, containing server's identity $ID_s$, a random seed $\emptyset$ and server's phone number $T_s$. The TSP then sends $ID_s$, $T_s$, $\emptyset$ and a shared secrete key $K_{sd}$ to the cell phone. After the response is received, the user can enter a long-term password $P_u$ in his cell phone.

The cell phone calculates a secret credential $C$ by the following operation:

$$C = H(P_u \| ID_s \| \emptyset) \qquad (1)$$

The cell phone encrypts the computed credential $C$ with the key $K_{sd}$ and generates the corresponding MAC, i.e., HMAC. HMAC-SHA1 obtains input user's identity, cipher text and IV to output the MAC [17], [18] and prepares this as a secure registration SMS. Then, the encrypted registration SMS is sent to the server by phone number $T_s$ as follows:

$$\text{Cell phone} \xrightarrow{\text{SMS}} \text{Server}\big(ID_u, \{C\|\emptyset\}_{K_{sd}}, IV, HMAC - SHA_1\big) \quad (2)$$

The authenticity of the registration SMS is checked by decrypting it and then information is obtained with the shared secrete key $K_{sd}$. To avoid SMS spoofing attacks, Server checks the source of received SMS with $T_u$. At last, the cell phone stores all information $\{ID_s, T_s, \emptyset, i\}$ excluding for the long-term password $P_u$ and the secret $C$. The current index of the one-time password is specified by variable $i$ and is initially set to 0. The server authenticates the user device during each login with $i$. On reception of the message (6), the server stores $\{ID_u, T_u, C, \emptyset, i\}$ and finishes the registration phase.

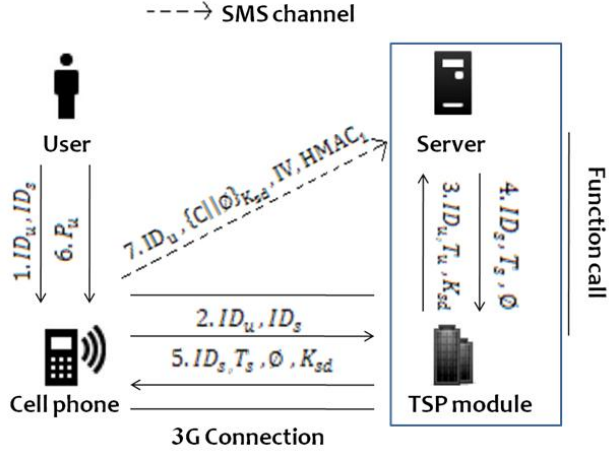

Fig. 1. Process of Registration Phase

## C. Login Phase

Fig. 2 shows the detail flow of the login phase. The user starts by sending login request to the server through an untrusted browser (on a kiosk) with his account $ID_u$. Then, server supplies the $ID_s$ and fresh nonce $n_s$ to the browser. In the meantime, this message is forwarded to the cell phone through Wireless or Bluetooth interfaces. After reception of the message, the cell phone checks related information from its database via $ID_s$ which comprises server's phone number $T_s$ and other parameters $\{\emptyset, i\}$. After proper inquiry, it shows a dialog to enter a long-term password $P_u$. Insert a correct long-term password $P_u$ on the cell phone to regenerate secrete shared credential $C$. The one-time password $\delta_i$ for current login is recomputed using the following operations:

$$C = H(P_u \| ID_s \| \emptyset) \quad (3)$$
$$\delta_i = H^{N \ldots i}(C) \quad (4)$$

$\delta_i$ is only used for this login (ith login after user registered) and for encryption used as a secret key with AES-CBC. The cell phone produces a fresh nonce $n_d$ and encrypts $n_d$ and $n_s$ with $\delta_i$ to generate the related MAC, i.e., HMAC and prepares a secure login SMS which is then sent to server S:

$$\text{Cell phone} \xrightarrow{\text{SMS}} \text{Server}\left(ID_u, \{n_d \| n_s\}_{\delta_i}, IV, HMAC - SHA_2\right) (5)$$

The server recomputed $\delta_i$ (i.e., $\delta_i = H^{N \ldots i}(C)$) after receiving the login SMS to decrypt and confirms the authenticity of the login SMS. If the nonce $n_s$ received equal the prior generated nonce $n_s$, the user is valid one; else, the server will reject the login request. The server sends a success message through the internet, $H(n_d \| \delta_i)$ to the user's cell phone by successfully verifying user's identity. The cell phone checks the received message to finish the login procedure. The last verification on the cell phone avoids the phishing attacks and the man-in-the-middle attacks. If verification fails, the user knows a reason and the device would not increment the index $i$. After successful user login, index will automatically increase, $i = i + 1$ simultaneously in both the device and the server for synchronization of one-time password. To refresh one-time password, user and the server can reset random seed $\emptyset$ by the recovery phase after specific period, $N - 1$ rounds (where N is pre-defined length of hash chain).

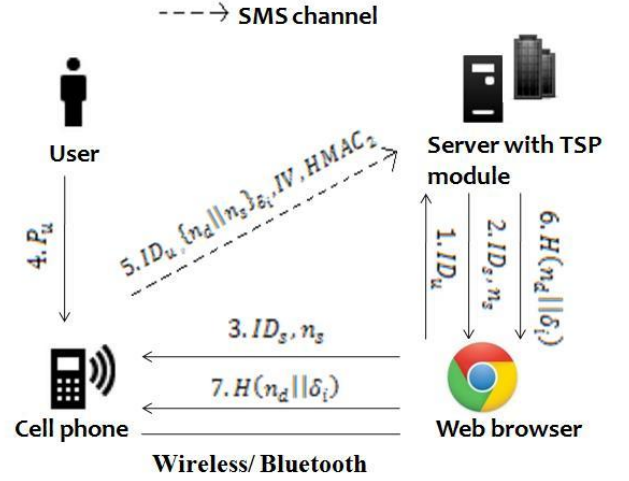

Fig. 2. Process of Login Phase

## D. Recovery Phase

Recovery phase is important and required in some particular conditions (e.g., a user may lose her cell phone.). The system can recover setting on her new cell phone (apply a new SIM card with old phone number (Number portability)). When user installs the program on new cell phone and launches the application, user asks for a recovery request with same previous account $ID_u$ and requested server $ID_s$ to TSP module through a 3G connection. As we stated earlier that $ID_s$ can be the domain name or URL link of server. Same in the registration process, TSP module can trace user's phone number $T_u$ based on her SIM card and sends user's account $ID_u$ and $T_u$ to server through a functional call. After getting the request, server searches the account information in its database to verify if account is previously registered or not. If account $ID_u$ is already there then the information used to calculate the secret credential will be extracted and be returned to the user. The server produces fresh nonce $n_s$ and replies with a message which contains $ID_s$, $\emptyset$, $T_s$, $i$ and $n_s$. This message contains all the required parameters to produce the next one-time passwords to the user.

Same in the registration process, the mobile application receives the message; it asks the user to enter his long-term password to regenerate the correct one-time password $\delta_{i+1}$ (presuming the last successful login before user lost his cell phone is $\delta_i$). Lastly, the user's cell phone produces a cipher text by encrypting the secret credential and server nonce $n_s$ and sent it back as a recovery message to the server for verification.

Likewise, the server calculates $\delta_{i+1}$ and decrypts this recovery message to confirm that user is already recovered. Now, user's new cell phone is recovered successfully and now ready to execute further logins. For the subsequent user login, one-time password $\delta_{i+2}$ will be used. Fig. 3 shows the detail flow of the recovery phase. Fig. 3 shows the detail flow of the recovery phase.

*E. Platform*

Platform: Windows (Windows 7, Windows XP)
Tools for programming: Android 2.2 SDK and its emulator must be installed, Eclipse IDE (versions 3.5.1 and higher), SQLite database, Apache server, MYSQL database.
Hardware: Processor-Intel Core2 Duo, RAM-1GB, Android device osv2.0 and above, GSM modem.
Technology: Java, Html, Xml, Android API, PHP, SMSLib (Open source library).

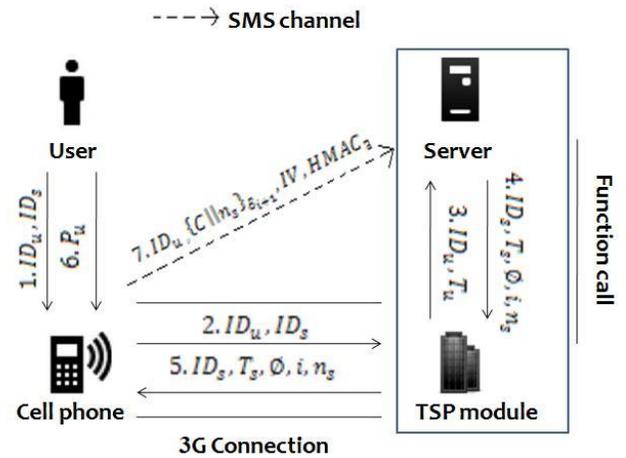

Fig. 3. Process of Recovery Phase

IV. RESULT ANALYSIS

Table I evaluates our system with earlier systems indicating avoided attacks. Symbol ' ✓ ' shows that the system avoids attacks, and ' − ' represents "not applicable". Table II compares our system with earlier systems for requirements. Symbol ' • ' indicates the special requisites, and ' − ' represents "not applicable".

Comparing our system with previous research

TABLE I

| System | Attack Prevention | | | | | |
|---|---|---|---|---|---|---|
| | Session hijacking | Phishing | Key-logging | Password reuse | DNS spoofing | Malware prevention |
| Our system | ✓ | ✓ | ✓ | ✓ | ✓ | ✓ |
| oPass [1] | ✓ | ✓ | ✓ | ✓ | ✓ | ✓ |
| MP-Auth [4] | ✓ | ✓ | ✓ | − | ✓ | ✓ |
| Phool Proof [2] | ✓ | ✓ | ✓ | | | ✓ |
| Secure Web[3] | ✓ | ✓ | ✓ | | ✓ | ✓ |
| BitE [15] | | | ✓ | | ✓ | ✓ |
| Garriss et al.[16] | | | ✓ | | | ✓ |
| Session Magnifier[14] | ✓ | | | | ✓ | |
| Secure Pass [5] | ✓ | ✓ | ✓ | ✓ | ✓ | ✓ |

TABLE II

UICC STANDS FOR USER INVOLVEMENT IN CERTIFICATE CONFIRMATION
TPM STANDS FOR TRUSTED PLATFORM MODULE

| System | Requirements | | | | | | |
|---|---|---|---|---|---|---|---|
| | UICC | Physical account setup | Logical account setup | TPM | On device secret | Trusted proxy | Malware free mobile |
| Our system | | | • | | | • | |
| oPass [1] | | | • | | | • | • |
| MP-Auth [4] | • | • | | | | | • |
| Phool Proof [2] | • | • | | | • | | • |
| Secure Web[3] | | • | | | • | • | • |
| BitE [15] | | | | • | • | | • |
| Garriss et al.[16] | • | | — | • | • | | • |
| Session Magnifier[14] | | | — | | • | | • |
| Secure Pass [5] | | • | — | | • | • | • |

## V. Conclusion and Future Scope

This paper proposes a novel architecture for secure user authentication avoiding possible threats. It uses the concept of two-factor authentication, one-time password and telecommunication services along with web services. This methodology of authentication is safe and reliable; provides high data confidentiality. Users need not to type any passwords into untrusted computers for login on all websites; without the secret code of the user account, no one can extract the data. System implements the one-time password approach to ensure independence between each login session. Compared with earlier methods, the user authentication protocol effectively prevents password stealing (i.e., phishing, key logger) and reuse attacks. Password recovery is also provided when users lose their cell phones to produce fully functional system. User can recover system with reissued SIM cards and long-term passwords. In current internet environment most of the users are familiar with systems are accessible by using two-factors (tokens). Hence, in the future work, we plan to implement a system works on multi-parameters to authenticate user e.g., three-factor authentication (Biometric) and more secure techniques for secure message transmission.